# Dynamic hysteresis of an oscillatory contact line

Jiaxing Shen[1], Yaerim Lee[1], Yuanzhe Li[2], Stéphane Zaleski[3,4], Gustav Amberg[5] and Junichiro Shiomi[1,2]

[1] Department of Mechanical Engineering, Graduate School of Engineering, The University of Tokyo, Hongo, Bunkyo-ku, Tokyo 113-8656, Japan.

[2] Institute of Engineering Innovation, School of Engineering, The University of Tokyo, 7-3-1 Hongo, Bunkyo-ku, Tokyo 113-8656, Japan.

[3] Sorbonne Université and CNRS, UMR 7190, Institut Jean Le Rond d'Alembert, 75005 Paris, France.

[4] Institut Universitaire de France, UMR 7190, Institut Jean Le Rond d'Alembert, 75005 Paris, France.

[5] Södertörn University, SE-141 89 Stockholm, Sweden.

**Abstract**

During oscillatory wetting, a phase retardation emerges between contact angle variation and contact line velocity, presenting as a hysteresis loop in their correlation—an effect we term dynamic hysteresis. This phenomenon is found to be tunable by modifying the surface with different molecular layers. A comparative analysis of dynamic hysteresis, static hysteresis, and contact line friction coefficients across diverse substrates reveals that dynamic hysteresis is not a result of dissipative effects but is instead proportionally linked to the flexibility of the grafted layer on the surface. In the quest for appropriate conditions to model oscillatory contact line motion, we identify the generalized Hocking's linear law and modified Generalized Navier Boundary Condition (GNBC) as alternative options for predicting realistic dynamic hysteresis.

## I. Introduction

The phenomenon of liquid spreading along a solid surface by displacing gas is termed dynamic wetting. This process occurs ubiquitously in nature, everyday experiences, and various technological applications such as coating and printing. The extensive relevance of wetting process highlights its practical significance. At the wetting front, three interfaces between each pair of the three materials intersect, forming a contact line region. When the influence of surface tension is significant, as indicated by a small capillary number (Ca), the dynamics of the contact line become important for the entire flow. This is because the contact angle, acting as a boundary condition, shapes the fluid interface [1–3]. In this context, the contact line dynamics emerges as the fundamental problem of wetting.

Understanding the physics of a moving contact line (MCL) is challenging owing to its multiscale features [4–6]. Usual optical measurements can capture the interface profile only at a resolution poorer than the microscale, leaving the crucial details at finer scales unresolved. The angle between the interface and the substrate, accessible through optical measurements, is known as the macroscopic or apparent angle $\theta_{app}$. During wetting or dewetting, $\theta_{app}$ deviates from its equilibrium value $\theta_e$, a deviation attributed to contributions from various scales. In a bottom-up manner, at the molecular level, the microscopic contact angle demonstrates velocity dependence owing to molecule jumping activities as explained by molecular kinetic theory (MKT) [7]. Beyond the molecular region is the nanobending region, a convex nanoscopic structure [6,8] that links the microscopic angle $\theta_m$ with the macroscale region. Chen *et al.* [8] recently revealed this mesoscopic link of advancing contact lines using tapping mode atomic force microscopy (AFM). The curvature of nanobending structure and $\theta_m$, the root of nanobending region, are both velocity-dependent. The mesoscopic angle, $\theta_{me}$, is defined at the end of the the nanobending region, which is measured to be 20 nm in height. Beyond this level, the Ca-dependent concave viscous bending becomes prominent, forming the main focus of hydrodynamic models [9,10]. The multiscale structure of the complex contact line is depicted in Figure 1.

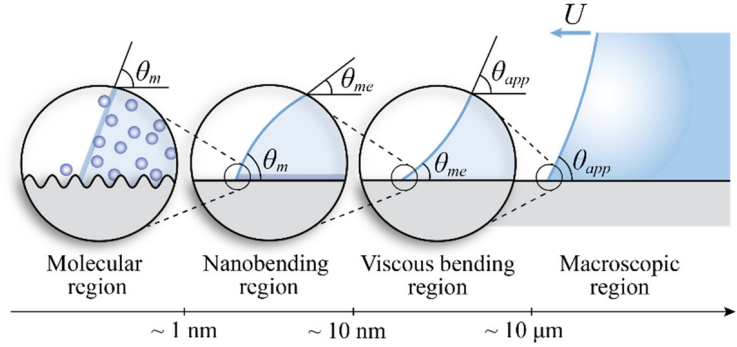

FIGURE 1. Sketch of the multiscale structure of contact angle during advancing.

In practical situations, the interest usually lies in the dependence of the apparent contact angle $\theta_{app}$ on contact line velocity $U_{CL}$, rather than its complex origins involving different dissipation channels at various scales, such as molecular friction and viscous dissipation. From another perspective, the relation between dynamic contact angle and contact line speed is also important for modelling macroscale flows with moving contact lines. Imposing a prescribed (velocity-dependent) interfacial angle as a boundary condition can address the microscopic effects on the macroscopic flow [11–13]. This approach relaxes the requirement to resolve nano-/micro-scale details near the contact line, making it an alternative to computationally expensive highly resolved directional numerical simulations [14,15]. These considerations drive our interest in exploring the constitutive relation between dynamic contact angle and contact line velocity in this study. Empirical observations commonly suggest a nonlinear connection between the macroscopic contact angle and velocity [1,16]. While in many cases Hocking's linear law [17], which relates the deviation of contact angle from equilibrium and the contact line velocity through a real number $\lambda$, is an adequate approximation to capture the dynamic contact line behavior:
$$U_{CL} = \lambda(\theta_d - \theta_e) \qquad (1.1)$$
where the slope $\lambda$ is termed different names in various studies, like capillary coefficient [18–20] and mobility [21,22]. By transforming the conventional $\Delta\theta$-$U_{CL}$ diagram of an oscillating wetting line into a $\Delta\theta\eta$-$U_{CL}\eta$ diagram, where $\Delta\theta = \theta_d - \theta_e$ represents the contact angle deviation from its equilibrium value and $\eta$ represents the contact line displacement, Xia and Steen [21] noted that the slope $\lambda$ remains a real constant in the regions away from the stick-slip and can function as a phenomenological parameter for evaluating contact line mobility.

However, the behavior of the contact line is influenced not only by the localized material properties of three-phase systems but also by the dynamics of the flow. It has been noted that the boundary condition for unsteady flow should differ from that of steady situations [19,20,23,24]. Hocking, in the study of the contact line problem for surface wave, implicitly considers the contact angle and contact line velocity to be in phase by assigning the capillary coefficient $\lambda$ as a real constant [17,18,23]. Miles [23] suggests that for unsteady contact line motion, the slope $\lambda$ in Hocking's linear law becomes a complex function of frequency, thereby introducing a phase offset between the contact angle and contact line velocity. This results in the formation of a hysteresis loop in the $\theta$-$U_{CL}$ diagram, as illustrated in figure 2 and also figure 4 of reference [20].

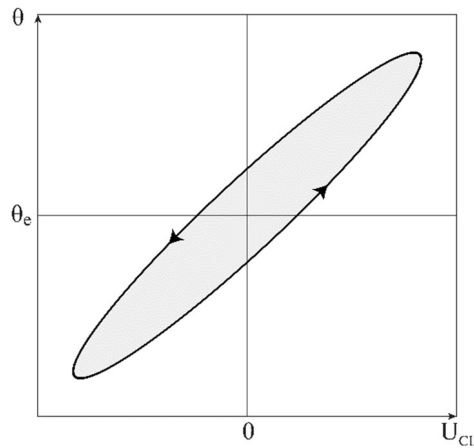

FIGURE 2. Illustration of the hysteresis loop in the $\theta$-$U_{CL}$ diagram predicted by Miles' model.

In the investigations of oscillating contact lines on vertical walls composed of different materials (glass and stainless steel), Perlin and the collaborators [19,20] observed a complex angle-velocity correlation resembling an inverted 'T' with a loop formed at the base. They experimentally evaluate the parameter $\lambda$, revealing its time-dependent nature. The openness of the angle-velocity curve is also observed in the contact line behavior of an oscillating drop on a fluorinated surface in the work by Xia and Steen [21] (supplementary material). However, Xia and Steen's analysis omits the hysteresis loop and attributes it simply to dissipation, which is contradicted by the results of the current study. Another example of the open hysteresis loop is documented in the molecular dynamics (MD) simulation conducted for an oscillating Wilhelmy plate experiment by Jin *et al.* [25]. They correlated the contact angle with the force exerted by the liquid on the solid wall and observed a hysteresis loop in the force-velocity diagram on a microscopically rough surface. In their study, this phenomenon is attributed to the broadening of the interface on the rough surface.

Upon the brief discussion, it is evident that the dynamic contact angle may not be in phase with contact line velocity in the case of an unsteady motion. Consequently, the contact line behavior deviates from that observed in steady motion, defying a description by a single-valued function connecting angle and velocity. To comprehend the dynamics of the contact line under unsteady conditions, this study explores the mobile contact line of a sessile droplet supported by a vertically vibrating substrate through experiment. Our observations reveal a distinct hysteresis loop in the oscillatory contact line behavior, which we term 'dynamic hysteresis,' distinguishing it from the static hysteresis defined as the interval between the critical advancing and receding angles. This dynamic hysteresis is related to the time retardation between the contact angle and contact line velocity. Through various surface molecule modifications, it becomes apparent that the dynamic hysteresis is sensitive to both static hysteresis and specific molecular properties (chain flexibility). Notably, we establish that the dynamic hysteresis remains unrelated to the magnitude of the contact line friction coefficient, indicating that it is not attributable to dissipative effects. Furthermore, we made attempts from various perspectives to determine suitable models for predicting the behavior of an oscillatory contact line. First, we generalize Hocking's linear law by making the coefficient λ a function of the time derivative of the logarithm of contact line velocity. Next, we assessed the dynamic hysteresis predicted by generalized Navier boundary condition (GNBC) and its derivatives, which intrinsically encompass the time derivative of contact angle due to the velocity gradient term.

This study explores the intriguing phenomena associated with oscillating contact lines. The molecular-scale interaction between the liquid and solid materials in the proximity of an oscillating contact line not only impacts the dissipation rate but also contributes to the memory effect on the dynamic contact angle. In this context, the constitutive law for the dynamic contact angle should incorporate acceleration in addition to contact line velocity.

## II. Experimental Method

### 2.1 Oscillatory wetting experiment

The test rig of the oscillatory wetting experiments is depicted in figure 3(a). In this configuration, a 10 μL pure water droplet is placed on substrate sample, which is affixed horizontally to a vertically vibrating stage. The vibration frequency is set at 70 Hz, close to the resonance frequency of [2, 0] axisymmetric mode [26], determined through a rough frequency sweep. To isolate and highlight the influence of substrate-related factors, we maintain a constant frequency. This aims to prevent the introduction of effects arising from the change of Stokes viscous layer, the thickness of which is denoted by $\delta = \sqrt{\mu/\rho\omega}$. Driven by the axisymmetric bulk motion, the contact line advances and recedes periodically, along with the dynamically changing contact angle (refer to figure 3(b)). The mechanical vibration system comprises a function generator (low-frequency power oscillator URP-20, SHIMADZU) and a mechanical vibrator (Mechanical wave driver SF-9324, PASCO scientific).

To capture the transient contact line movement, a high-speed camera (Phantom VEO710L, Vision Research Inc.) equipped with a Tamron SP AF 180mm F/3.5 Di macro lens (Tamron USA, Inc., Commack, NY) is employed, illuminated by a backlight lamp (HVC-SL, Photron). The frame rate is set at 7000 fps. The high-speed visualization system is enhanced with a 5x objective lens (OLYMPUS LMPLFLN), improving the spatial resolution to 3 μm/pixel.

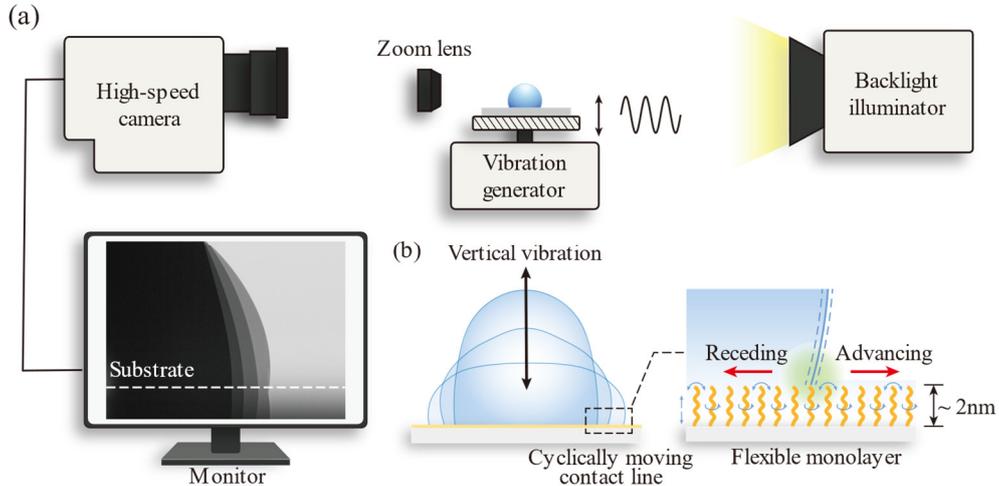

FIGURE 3. (a) Test of the oscillatory wetting experiment. (b) Illustration of the local contact line movement on a monolayer modified substrate.

2.2 Image analysis

To automatically extract the dynamic contact angle and contact line position from the high-speed video frame by frame, we have developed a customized MATLAB program based on polynomial fitting approach. Here we provide a brief overview of the three main steps of our program; for detailed information, refer to the supplementary material.

1. Extracting drop boundary using Canny edge detector and obtaining a set of pixel coordinate representing drop boundary.

2. Identifying the vertical and horizontal contact line coordinates by employing the profile method [27].

3. Approximating the drop profile using a fourth order polynomial curve fitted with 200 pixels along the drop boundary above the contact point within the polar coordinate system [28]. Then the contact angle is estimated as the derivative of the fitted polynomial at the contact point.

Additionally, the contact line velocity is computed as the time derivative of the contact line's horizontal position using central differencing.

2.3 Substrate preparation

Silicon wafers were first cut into 2 cm × 2 cm chips to serve as substrates. Prior to surface modifications, silicon chips were sequentially cleaned in ultrasonic baths of acetone, ethanol, distilled water for 5 min each and then dried with nitrogen gas. Following the cleaning process, a 15-min ozone plasma treatment was applied to the samples to remove organic contaminants and to enhance the adhesive property of the surfaces.

In our experiments, four kinds of molecules were used for modification of the substrates: Trichloro(3,3,4,4,5,5,6,6,7,7,8,8,8-tridecafluorooctyl)silane (FUJIFILM Wako Pure Chemical Corporation, JP), referred to as fluoroalkyl silane (FAS), trimethylsiloxy terminated polydimethylsiloxane (PDMS, M.W. 2000) (Thermo Fisher Scientific, USA), and two alkyl silanes with distinct lengths—trimethoxy-n-octylsilane (AS-C8) (FUJIFILM Wako Pure Chemical Corporation, JP) and octadecyltrimethoxysilane (AS-C18) (FUJIFILM Wako Pure Chemical Corporation, JP). These molecules are grafted to silicon chips through different procedures according to different chemisorption mechanisms.

Fluoroalkyl silane was grafted onto silicon using the vapor phase deposition approach. The silicon chip and vaporized fluoroalkyl silane reacted in a vacuumized desiccator at room temperature for 2 hours, ensuring the substrate to reach a state of saturated hydrophobicity.

The alkyl silanes were coated on silicon through an immersion technique. The silicon chip was immersed in a toluene solution containing 0.5 mM/L alkyl silane at room temperature for 18 h, catalyzed by HCl. After the reaction, the surfaces were rinsed by toluene, ethanol, and distilled water in order.

The PDMS is covalently attached to silicon substrate by heat treatment. The silicon chip was wetted by the undiluted PDMS melt in a capped vial and subsequently baked in an oven at 100 °C for 24 h. Following the reaction, the surfaces were rinsed in a sequential order using toluene, ethanol, and distilled water.

Lastly, bare silicon substrates without a monolayer were included for comparative analysis. In this case, the silicon chip was immersed in a 50% HF solution for 10 min to remove the native oxide layer. This treatment increases the static contact angle of bare substrate, which is preferred in our zoomed visualization system.

All the substrate samples used in the drop oscillation experiments were freshly made within 2 days to avoid aging effect of the grafted molecule layer.

### 2.4 Surface characterization

The static contact angle and contact angle hysteresis (CAH) were assessed using a high-precision automatic contact angle meter (DropMaster, model DMo-602, Kyowa). The equilibrium contact angle $\theta_e$ is measured by circular fitting of the drop's contour (axisymmetric), captured from a side view after gently depositing a 10 μL water droplet on the targeted sample. The CAH is measured through tilt plate method [29] with the same droplet. The thickness of the monolayer grafted onto the silicon chip was determined using a spectroscopic ellipsometer (M-2000U, J.A. Woollam). The static wetting properties and monolayer thickness of various surfaces are summarized in Table 1.

TABLE 1. Wetting properties and monolayer thickness of various surfaces.

| Substrates | $\theta_e$ | S.D. | $\theta_{adv}$ | S.D. | $\theta_{rec}$ | S.D. | $\Delta\theta$ | S.D. | Thickness (nm) |
|---|---|---|---|---|---|---|---|---|---|
| FAS | 105° | 0.8° | 116° | 1.0° | 95° | 2.7° | 21° | 2.2° | 1.55 |
| AS-C8 | 105° | 1.3° | 108° | 1.0° | 97° | 1.2° | 11° | 1.5° | 0.94 |
| AS-C18 | 104° | 0.4° | 107° | 2.2° | 98° | 0.3° | 10° | 2.3° | 1.97 |
| PDMS | 104° | 0.7° | 106° | 0.8° | 99° | 1.5° | 7° | 1.4° | 2.03 |
| Silicon | 69° | 1.1° | 77° | 4.7° | 50° | 2.0° | 27° | 2.6° | — |

**III. Results**

Conventionally, the behavior of contact line is described by a single-valued dependence of the dynamic contact angle $\theta_d$ on the contact line velocity $U_{CL}$. However, by performing oscillatory wetting experiments on substrates modified by various molecules, a distinct hysteresis loop has been identified in the angle-velocity correlation, indicating the impact of memory effect on the oscillatory contact line movement.

### 3.1 Hysteresis loop

To study the contact line behavior in a holistic way, we graph the trajectory of the cyclic contact line movement in a 3D phase space. The three dimensions correspond to contact line displacement, contact line velocity, and dynamic contact angle, respectively, as shown in figure 4(a). The circulation occurs in a clockwise direction when viewed from top to bottom. The primary distinction in the phase trajectories of contact line motion on silicon and PDMS-coated surfaces is evident in their projection on the angle-velocity plane. On the PDMS-modified surface, the correlation between $\theta_d$ and $U_{CL}$ is nearly single-valued, resembling the conventional contact line relation observed in unidirectional motion [1]. In contrast, the angle-velocity relation is more complex on bare silicon substrate, presenting as a hysteresis loop. Similarly, the dynamic hysteresis is observed on fluoroalkyl silane and alkyl silane modified surfaces, although the loop exhibits distinct features across different surfaces, see figure 4(b).

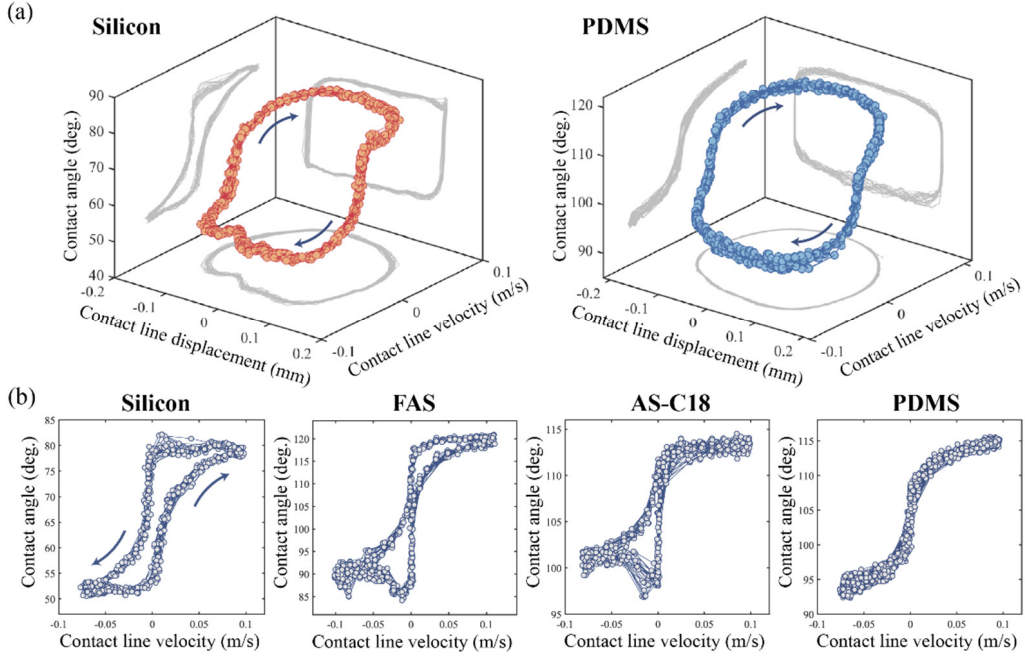

FIGURE 4. (a) 3d plotting of phase planes consisting of contact angle, contact line velocity and contact line position. (b) Angle-velocity relation of different surfaces. The arrows in the figures indicate the system's evolution over time.

To quantitatively compare the dynamic hysteresis across different substrates, we characterize the hysteresis loop by calculating the ratio between the area enclosed by the angle-velocity curve and the rectangular area enclosed by the four extrema. This is illustrated in the inset of Figure 5(b). In this manner, we have determined that the dynamic hysteresis depends on the surface material and is not influenced by the bulk flow. This is clearly illustrated in Figure 5(b), where the x-axis represents the peak acceleration of substrate vertical vibration, which is calculated as:

$$a_Y = (2\pi f)^2 Y_{amp}$$

By tuning the amplitude of plate vibration $Y_{amp}$, the flow condition is adjusted in response to different plate accelerations. There is no obvious correlation between the dynamic hysteresis and the plate acceleration, indicating that the dynamic hysteresis is not flow dependent. Meanwhile, dynamic hysteresis systematically varies across different substrates. The PDMS-coated surface exhibits the lowest dynamic hysteresis, approaching almost a single-valued curve. In contrast, the bare silicon chip displays the largest dynamic hysteresis in the angle-velocity diagram, while fluorinated alkyl silane and alkyl silane result in a moderate level of dynamic hysteresis, falling between PDMS and silicon.

Figure 5(a) depicts the static wetting properties of various substrates. A comparison between figure 5(a) and (b) reveals that the dynamic hysteresis of tested surfaces mirrors a similar trend to the static hysteresis. The ranking is as follows: silicon > fluoroalkyl silane > alkyl silane > PDMS. The ranking of static hysteresis to some extent reflects the flexibility of the surfaces. The brush-like PDMS is highly flexible, rendering a liquid-like property to the surface and resulting in the lowest static contact angle hysteresis [30,31]. In contrast, the chain lengths of alkyl silane and fluoroalkyl silane in our experiments are significantly shorter than the PDMS brush, leading to reduced chain mobility in the grafted layer [31]. Additionally, the fluorocarbon chain is inherently "stiffer" than its hydrocarbon counterpart [32]. Finally, the bare silicon chip can be considered inflexible, given the presence of only nanoscale solid asperities on the surface, consequently leading to the highest contact angle hysteresis. However, the relationship asserted here between the hysteresis and the flexibility of the surface layer is qualitative. The rigorous validation of this trend through complex characterization is beyond the scope of this study.

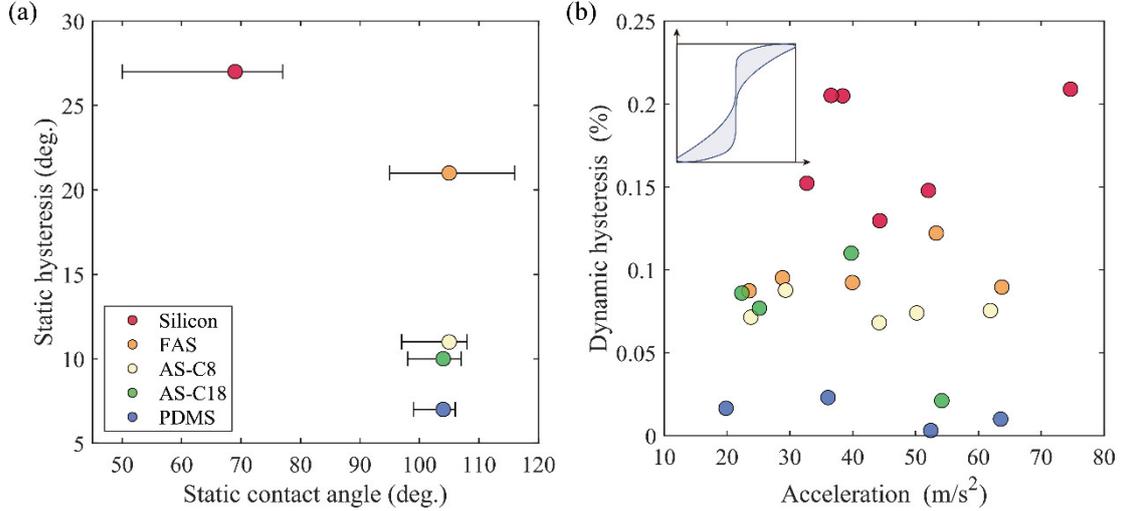

FIGURE 5. (a) Static wetting properties of different substrates; (b) Quantified dynamic hysteresis of different substrates. Inset: Illustration of dynamic hysteresis quantification using the normalized area.

### 3.2 Phase difference

Figure 6 presents the time histories of contact line velocity and dynamic contact angle for two representative cases, namely bare-silicon and PDMS-coated surfaces. The comparison emphasizes a phase delay of the velocity signal to the contact angle on the silicon surface. This phase difference can be expressed as:

$$\theta_d(t) = f[U(t - \tau)]. \tag{3.1}$$

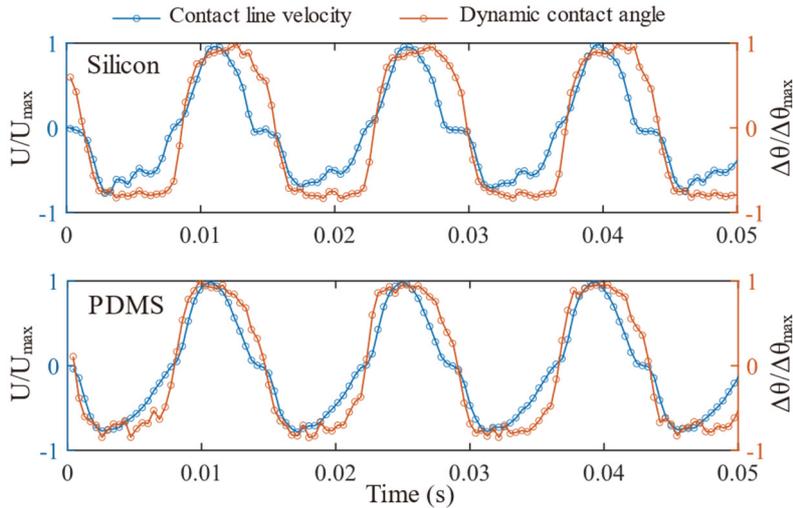

FIGURE 6. Time evolutions of contact angle and contact line velocity on Silicon substrate (top) and PDMS coated substrate (bottom).

It is crucial to exercise caution when comparing the direct time history of different variables,

given that the shapes of these signals are non-standard and do not conform to typical sinusoidal or square wave patterns. An approximate analysis may result in the loss of phase information. In a study by Cocciaro et al. [18], the contact angle signal was approximated as a square wave when investigating the contact line effect on standing surface waves. Their findings indicate a $\pi/2$ phase difference between the contact angle and contact line displacement signals, leading to the conclusion that the contact angle is in phase with the contact line velocity. Based on this observation, they suggest that the capillary parameter $\lambda$ in Hocking's linear law should be treated as a real number.

For this consideration, we take an alternative approach to evaluate the phase difference between the velocity and angle signals. We found that the hysteresis loop can be closed by shifting the velocity signal forward for a duration $\tau$, establishing that the contact angle $\theta_d$ is in phase with the variable $(U - \tau \dot{U})$, where $\dot{U}$ represents the contact line acceleration obtained through the time derivative of $U$ using central differencing. In this context, the dynamic contact line behavior can now be described by a single-valued function:

$$\theta_d = f[U(t - \tau)] = f(U - \tau \dot{U}) \tag{3.2}$$

The closure of the hysteresis loop is illustrated in figure 7(a).

Note that Ting and Perlin [19] also acknowledged the contribution of contact line acceleration to the dynamic contact angle, which, however, is considered in the sense that the acceleration is the derivative of velocity. In addition, in concerns of calculation error, they opted not to conclusively establish the relationship between the dynamic contact angle and contact line acceleration. In Eq. (3.2), we illustrate that the acceleration $\dot{U}$ contributes to $\theta_d$ owing to the retardation between angle and velocity. The relationship between $\theta_d$ and $\dot{U}$ extracted in our experiments is depicted in figure S3, exhibiting a 'Z' shaped loop among various surfaces.

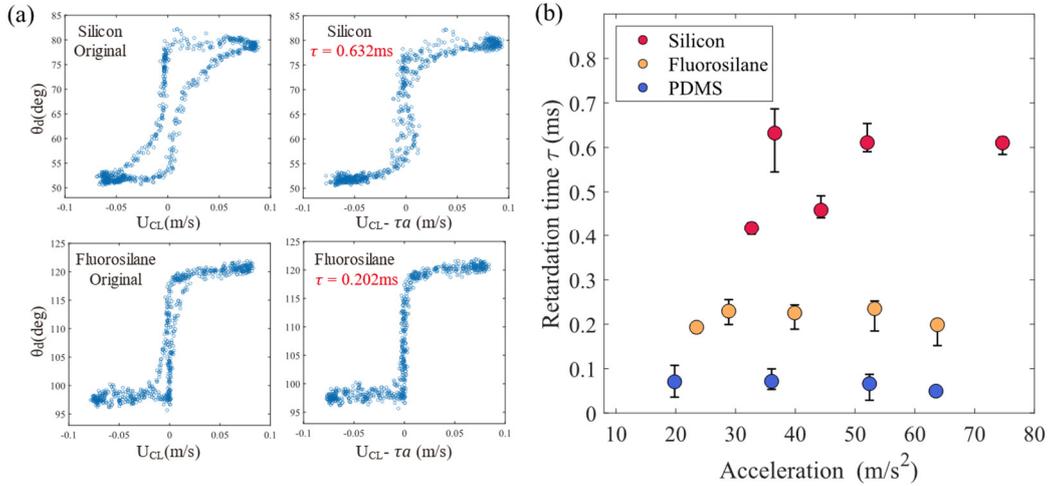

FIGURE 7. (a) Examples showing the closure of the hysteresis loop by shifting the velocity signal forward for a duration τ. Top: Silicon substrate; bottom: Fluoroalkyl silane coated substrate. (b) Retardation time τ of different substrates at different substrate accelerations.

The determination of $\tau$ involves a search algorithm conducted iteratively as follows: $\tau$ values are systematically calculated within a roughly estimated range with intervals of 0.001ms. Ultimately, the optimal retardation value, $\tau$, is selected based on the minimum absolute area enclosed by the curve of $(U - \tau \dot{U})$ versus $\theta$.

Figure 7(b) presents a summary of the extracted $\tau$ values across various surfaces and under different substrate accelerations. Two key observations emerge: firstly, retardation appears insensitive to substrate acceleration. Preliminary tests using different frequencies and different droplet volumes indicate similarly that $\tau$ is insensitive also to frequency variations (not presented in this paper). We conclude in this context $\tau$ can be treated as a material parameter. Secondly, $\tau$ values vary across substrates. The solid silicon surface exhibits the longest retardation time, whereas the PDMS-modified surface, with its fluid-like characteristics, yields the shortest retardation.

### 3.3 Contact line friction

The experimentally extracted angle is the macroscopic angle $\theta_{app}$, and its variation with contact line velocity is influenced by both viscous friction in the viscous bending region and local frictional dissipation at molecular scale. Instead of distinguishing between different dissipation channels, calculating the total dissipation is of practical interest. The total dissipation rate can be determined by considering that dissipation during contact line movement is entirely attributed to effective contact line friction. In this context, we introduce a method for experimentally deriving the contact line friction coefficient through cyclic contact line movement.

During one cycle of the cyclic motion, the total work done by the unbalanced Young's force is:

$$D_Y = 2\pi \int \gamma(\cos\theta_e - \cos\theta_d)\, r\, dr. \tag{3.3}$$

The dissipation at the contact line is accounted for in the form of friction:

$$D_f = 2\pi \int \mu_f U_{cl}\, r\, dr. \tag{3.4}$$

Given the prominence of contact line dissipation as the primary contributor to total dissipation [33], the integrated form of $\mu_{f,int}$ can be deduced by equating the total mechanical work to the frictional dissipation. Meanwhile, assuming a constant contact line friction coefficient and factoring it out from the integrand, the expression of $\mu_{f,int}$ is:

$$\mu_{f,int} = \frac{D_Y}{2\pi \int U_{cl}\, r\, dr}. \tag{3.5}$$

This can be calculated as the ratio of the area enclosed by the square in the left panel of figure 7 to the area enclosed by the circle in the right panel of the same figure. Utilizing the same energy balance relation in a discrete manner, the calculation of the discrete contact line friction coefficient is expressed as:

$$\mu_{f,i} = \frac{\Delta D_{Y,i}}{2\pi U_{cl,i} r_i \Delta r_i}, \tag{3.6}$$

where $\Delta D_{Y,i}$ represents the mechanical work done by the unbalanced Young's force within a discrete step:

$$\Delta D_{Y,i} = 2\pi\gamma(\cos\theta_e - \cos\theta_{d,i}) r_i \Delta r_i, \tag{3.7}$$

and the discrete frictional dissipation is:

$$\Delta D_{f,i} = 2\pi \mu_{f,i} U_{cl,i} r_i \Delta r_i. \tag{3.8}$$

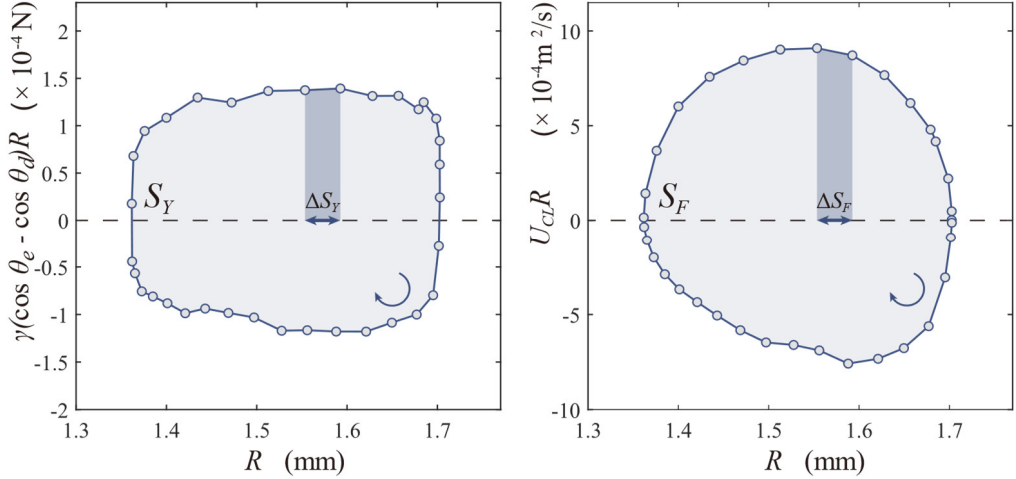

FIGURE 8. Evaluation of contact line friction coefficient.

The integrated form and discrete form of contact line friction coefficient $\mu_f$ are not standing along, but they can be related through the total work done by the uncompensated Young's force over a complete period:

$$\mu_{f,int} \cdot \left(\sum_{i=1}^{N} U_{cl,i} r_i \Delta r_i\right) = \sum_{i=1}^{N} \mu_{f,i} U_{cl,i} r_i \Delta r_i. \tag{3.9}$$

Therefore,

$$\mu_{f,int} = \frac{\sum_{i=1}^{N} \mu_{f,i} U_{cl,i} r_i \Delta r_i}{\sum_{i=1}^{N} U_{cl,i} r_i \Delta r_i} = \sum_{i=1}^{N} \left(\frac{U_{cl,i} r_i \Delta r_i}{\sum_{i=1}^{N} U_{cl,i} r_i \Delta r_i}\right) \mu_{f,i} = \sum_{i=1}^{N} \frac{\Delta S_{F,i}}{S_F} \mu_{f,i}. \tag{3.10}$$

According to Eq. (3.10), the integrated $\mu_{f,int}$ is a weighted average of discrete $\mu_{f,i}$, where the weight corresponds to the ratio of frictional dissipation in one step to that in a whole period. During a single discrete step, the discrete contact line friction coefficient can be treated as a constant. The cycle-averaged friction coefficient $\mu_{f,int}$ provides an overall estimation of the dissipation characteristics within a three-phase contact line system. In conjunction, the discrete coefficient $\mu_{f,i}$ offers instantaneous information about the resistance encountered during contact line motion. The experimental evidence presented in appendix A figure 11 emphasizes the dynamic nature of $\mu_{f,i}$.

Figure 9 presents the integrated contact line friction coefficient across various surfaces. In figure 9(a) a noticeable flow-dependent trend is observed as the integrated $\mu_{f,int}$ decreases with increasing substrate acceleration. From figure 9(b), we note a lack of correlation between the magnitude of contact line friction and dynamic hysteresis. This challenges the assumption that the openness of the hysteresis curve is attributable to dissipative effects [21]. Noteworthy differences emerge among surfaces, with the bare silicon surface exhibiting the highest dynamic hysteresis and contact line friction, while the long-chain alkyl silane (C18) demonstrates the lowest friction coefficient. PDMS and fluoroalkyl silane share a similar magnitude of contact line friction, yet PDMS showcases an almost single-valued angle-velocity relationship with minimal dynamic hysteresis, while fluorosilane's dynamic hysteresis ranks second only to the bare silicon substrate.

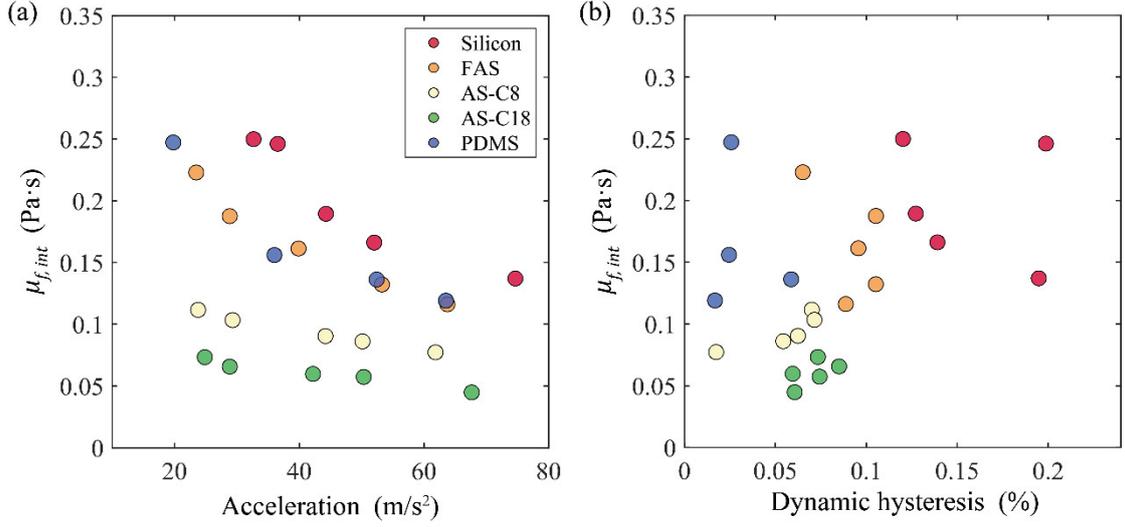

FIGURE 9. (a) Integrated $\mu_f$ on different substrates at different acceleration. (b) Integrated $\mu_f$ shows no obvious correlation with dynamic hysteresis.

Given the multiscale nature of the contact line and the diverse kinetics involved, various models account for the localized contact line dissipation using parameters resembling friction coefficients, including the line friction coefficient in diffuse interface modeling [34,35], the molecular line friction introduced in MKT [36], and the phenomenological parameter in Hocking's linear law [17,21], which is inversely proportional to the contact line friction. To prevent potential confusion regarding the experimentally extracted contact line friction coefficient, we provide a concise discussion in appendix B, outlining its relationship with other sources of contact line dissipation.

Moreover, while contact line friction is a primary focus of this study, one should keep in mind that it is not the sole method for describing dynamic wetting phenomena. Hydrodynamic model, which attributes the variation of dynamic contact angle with contact line speed purely to viscous bending while neglecting the molecular resistance near the contact line, have demonstrated success in predicting the wetting failure in curtain coating systems through highly resolved direct numerical simulations [14,15]. This lends support to the validity of hydrodynamic models and their assumption that the microscopic angle can be treated as constant in certain configurations.

### IV. Discussion

Dynamic hysteresis signifies a temporal misalignment between the dynamic contact angle and the velocity of the contact line. From a dynamic perspective, in oscillatory wetting, the free surface profile cannot be derived from a balance between pressure gradient and viscous shear stress as it can be under steady-state conditions [9], because in this case, the unsteady inertia term in the momentum equation becomes critical and cannot be omitted. Therefore, an acceleration dependence is introduced in the bending structure, leading to an expected retardation between angle and velocity. We extracted the acceleration dependence of dynamic contact angle from our experimental results, presented in figure SM3 (supplementary material). Despite similarities in this acceleration dependence, dynamic hysteresis varies significantly across different samples; for example, it is smallest on PDMS, largest on bare silicon. This substantial variation across surfaces underscores the importance of the liquid-solid interactions in this process and emphasizes the critical role of boundary conditions in modeling to tune the dynamic hysteresis.

Therefore, our primary focus remains on finding an appropriate constitutive relation that predicts the retardation by correlating the dynamic contact angle, CL velocity, and the time derivative of either one of them. This constitutive relation may serve as a convenient edge condition in moving contact line problems, providing an alternative to computationally expensive highly resolved DNS.

### 4.1 Generalization of Hocking's linear law

From the previous discussion on the oscillatory contact line motion, we found that the contact angle $\theta$ is in phase with the shifted velocity $U - \tau \dot{U}$. Consequently, we incorporate the shifted velocity into the Hocking's linear law Eq. (1.1):

$$(U - \tau \dot{U}) = \lambda_0 (\theta - \theta_e), \tag{4.1}$$

where $\lambda_0$ is a real number, not necessarily a constant. We rewrite (4.1) as:

$$U = \frac{\lambda_0}{\left(1 - \tau \frac{\dot{U}}{U}\right)} (\theta - \theta_e). \tag{4.2}$$

Comparing this relation with the original form of Hocking's linear relation, the generalized capillary coefficient is expressed as:

$$\lambda = \frac{\lambda_0}{\left(1 - \tau \frac{\dot{U}}{U}\right)} = \frac{\lambda_0}{\left(1 - \tau (\ln \dot{U})\right)}. \tag{4.3}$$

For sinusoidal contact line movement, if $U = U_0 e^{i\omega t}$, we can derive:

$$\lambda = \lambda_0 (1 + i\omega \tau). \tag{4.4}$$

This aligns with Miles's prediction [23], indicating that the coefficient $\lambda$ in Hocking's linear law is a function of the frequency $\omega$.

### 4.2 Retardation predicted by generalized Navier boundary condition

The boundary condition for a moving contact line typically involves the Navier boundary condition (NBC) to remove stress singularity, along with a constitutive relation that defines the contact angle variation with velocity. Constitutive relations, predicted by various theories such as the Cox-Voinov law [9,10], MKT-based model [36], or empirical relation like Kistler model [16], consistently involve only two variables—angle and velocity. However, the absence of the time derivative of either variable in these relations results in a prediction that the contact angle variation is always in phase with the contact line velocity. Consequently, these relations are not suitable for capturing oscillatory contact line motion.

However, the generalized Navier boundary condition (GNBC), which serve as an alternative to conventional contact line boundary condition, implicitly includes the time derivative of contact angle $\dot{\theta}$ in the shear term. This feature should result in a phase difference between the contact angle and the contact line velocity, giving rise to dynamic hysteresis. The GNBC describes that in the immediate vicinity of the contact line, the relative velocity of contact line is proportional to the summation of tangential stresses, encompassing the viscous shear stress and the unbalanced Young's stress [37]. The expression is as follows:

$$\beta U = \tau^V + \tau^Y. \tag{4.5}$$

Here, the slip coefficient $\beta = l_s/\mu$, with $l_s$ being the slip length and $\mu$ the liquid viscosity. The viscous stress $\tau^V = \mu\, \partial U/\partial n$, where $\partial/\partial n$ denotes the spatial derivative perpendicular to the wall. The integral of the uncompensated Young's stress $\tau^Y$ over the fluid-fluid interfacial region is the unbalanced Young's force [37]:

$$\int_{interface} \tau^Y\, dx = \gamma(\cos\theta_e - \cos\theta). \tag{4.6}$$

According to the MD simulation, the distribution of $\tau^Y$ along the flat substrate (parallel to the $x$ direction) is concentrated near the interfacial region, which extends about $10\sigma$ [37]. In MD simulations, $\sigma$ represents the range of interaction, typically around a few angstroms for many substances. Based on this, we approximate $\tau^Y$ to be uniformly distributed over a thickness $\xi \approx 5nm$ around the fluid-fluid interface. Thus, the GNBC is given by:

$$U = \frac{l_s}{\mu}\left[\mu\frac{\partial U}{\partial n} + \frac{\gamma}{\xi}(\cos\theta_e - \cos\theta)\right]. \tag{4.7}$$

Through a kinematic approach, Fricke et al. [38] derived that the time derivative of the dynamic contact angle $\dot\theta$ can be expressed in terms of the velocity gradient at the solid wall $\partial U/\partial n$. Thus, the relation can be rewritten as:

$$U = l_s\dot\theta + \frac{l_s\gamma}{\xi\mu}\sin\theta_e\,(\theta - \theta_e). \tag{4.8}$$

To simplify the problem, we model both the contact angle and contact line velocity as sinusoidal signals, expressed as:

$$\theta - \theta_e = \Delta\theta e^{i\omega t}, \tag{4.9}$$
$$U = U_0 e^{i\omega t}. \tag{4.10}$$

In these relations only the real part has physical meaning. Substituting these expressions into GNBC:

$$U_0 - il_s\omega\Delta\theta = \frac{l_s\gamma\sin\theta_e}{\xi\mu}\Delta\theta. \tag{4.11}$$

Thus

$$\Delta\theta = \frac{U_0}{\frac{l_s\gamma}{\xi\mu}\sin\theta_e + il_s\omega}. \tag{4.12}$$

Multiplying both sides of (4.12) by $e^{i\omega t}$ and substituting it back into (4.9):

$$\theta = \theta_e + \frac{U}{\frac{l_s\gamma}{\xi\mu}\sin\theta_e + il_s\omega} = \theta_e + \frac{U_0 e^{i\omega\left(t+\frac{\alpha}{\omega}\right)}}{\sqrt{\left(\frac{l_s\gamma}{\xi\mu}\sin\theta_e\right)^2 + (l_s\omega)^2}}, \tag{4.13}$$

where $\tan\alpha = -\frac{\omega}{\frac{\gamma}{\xi\mu}\sin\theta_e}$. Thus, the retardation time predicted by GNBC is:

$$\tau = -\frac{\alpha}{\omega} = -\frac{1}{\omega}atan\left(-\frac{\omega}{\frac{\gamma}{\xi\mu}\sin\theta_e}\right). \tag{4.14}$$

In the above calculation, by utilizing parameter values from our experiment— $\omega = 2\pi f$, $f = 70Hz$, $\gamma = 0.072N/m$, and considering the liquid properties of water: $\mu = 0.001Pa \cdot s$, $\sin\theta_e = 0.97$— the retardation time $\tau$ is calculated to be $0.07ns$.

However, this prediction made by GNBC contradicts experimental observations in two aspects. First, the retardation time predicted by GNBC is much shorter than the experimental results, which is on the order of $0.1ms$, as evidenced in figure 7(b). The second is that surfaces with close wettability ($\theta_e$), such as PDMS and Fluoroalkyl silane, exhibit different retardation times in experiments (see figure 7(b)). However, this difference, which arises from surface properties other than wettability, cannot be captured by GNBC prediction. This limitation is evident in (4.14), where only the wettability of the surface is involved.

From this discussion, it is evident that while GNBC can predict a dynamic hysteresis by introducing phase difference between $\theta$ and $U$, the predictions deviate from experimental observations.

### 4.3 Retardation predicted by modified GNBC

The failure identified in GNBC does not negate the potential for predicting dynamic hysteresis using this approach. However, to enhance its predictive capability, we need to incorporate additional surface properties into this relation. Consequently, we modify the GNBC by defining the slip velocity to be proportional to weighted summation of the two tangential stresses, where the weight of the unbalanced Young's stress is inversely proportional to the contact line friction coefficient $\mu_f$. The modified GNBC is expressed as:

$$U = l_s \frac{\partial U}{\partial n} + \frac{\gamma}{\mu_f}(\cos\theta_e - \cos\theta). \tag{4.15}$$

Again, rewrite the shear $\partial U/\partial n$ as $\dot{\theta}$:

$$U = l_s \dot{\theta} + \frac{\gamma}{\mu_f}\sin\theta_e \, (\theta - \theta_e). \tag{4.16}$$

Using the same treatment as in the last section we can derive:

$$\theta = \theta_e + \frac{U}{\frac{\gamma}{\mu_f}\sin\theta_e + i l_s \omega} = \theta_e + \frac{U_0 e^{i\omega\left(t+\frac{\alpha}{\omega}\right)}}{\sqrt{\left(\frac{\gamma}{\mu_f}\sin\theta_e\right)^2 + (l_s\omega)^2}}, \tag{4.17}$$

$\tan\alpha = -\frac{l_s\omega}{\frac{\gamma}{\mu_f}\sin\theta_e}$. So the retardation time is expressed as:

$$\tau = -\frac{1}{\omega}\,atan\left(-\frac{l_s\omega}{\frac{\gamma}{\mu_f}\sin\theta_e}\right). \tag{4.18}$$

In (4.18), there are three undetermined parameters: $\tau$, $\mu_f$, and $l_s$. By employing the typical values of $\tau$ and $\mu_f$ obtained from experiment, we can utilize Eq. (4.18) to estimate the slip length $l_s$:

$$l_s = -\tan(-\tau\omega)\frac{\gamma}{\omega\mu_f}\sin\theta_e.$$

Figure 10(b) presents the slip lengths obtained through this approach for the various tested substrates and accelerations. These values are much larger than the slip length typically measured, which is no more than a few hundred nanometers [39–41], yet for now we temporarily treat it as a fitting parameter.

The modified GNBC offers the possibility of defining contact line conditions for oscillatory wetting, enabling more realistic predictions of retardation time with the appropriate selection of slip length and contact line friction coefficient. In figure 10(a), the curve predicted by the modified GNBC is generated by inputting the shifted velocity signal (directly extracted from the experiment) $U(t - \tau)$ into eq (4.17) to obtain $\theta(t)$, which is then plotted against $U(t)$. Despite the hysteresis loop's comparable area to that obtained from the experiment, it is evident from figure 10(a) that when $\mu_f$ is treated as a constant, the shape still differs between the prediction and experimental results.

To enhance the predictive accuracy of the modified GNBC, we can further refine the model by considering $\mu_f$ as a function of the contact angle. This function may resemble the one presented in figure A1(c), or given by ref [33].

However, a notable limitation of this modified GNBC becomes apparent when considering slip lengths across different substrates as derived from this model. Figure 10(b) shows that the most hydrophilic silicon substrates exhibit the largest slip lengths compared to other hydrophobic surfaces. This result contradicts the expected trend where slip length typically increases with hydrophobicity [41,42]. Such discrepancies suggest potential shortcomings in the modified GNBC formulation.

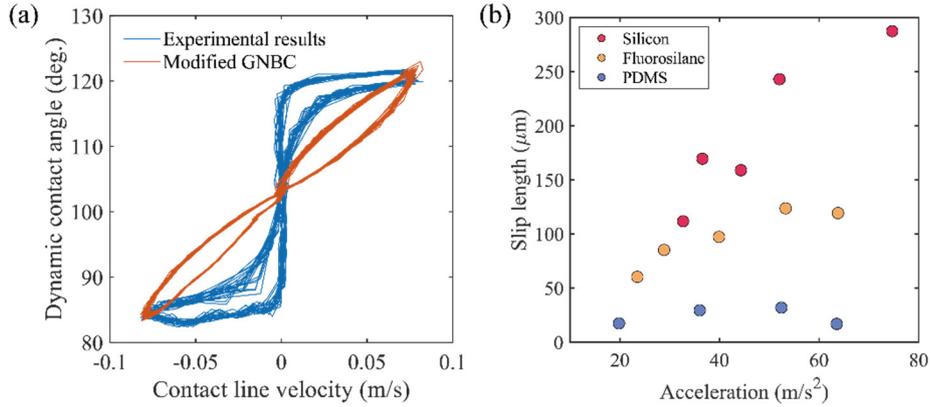

FIGURE 10. (a) Comparison between the angle-velocity relation obtained from experiment and predicted by modified GNBC. (b) Slip lengths extracted according to the modified GNBC for various substrates and accelerations.

## IV. Conclusion

The dynamic hysteresis of oscillatory contact line behavior is experimentally investigated in this work, manifesting as a hysteresis loop in the angle-velocity diagram. Our results underscore the necessity for a constitutive relation tailored to unsteady contact line motion, distinct from that derived under steady conditions. Molecular modifications on the surface induce variations in surface properties beyond wettability. A comparative analysis of oscillatory contact line dynamics on surfaces grafted with PDMS, fluoroalkyl silane, alkyl silane, and a bare silicon chip for reference reveals distinct dynamic hysteresis patterns, attributed to differing flexibilities of the surface layers. Notably, the observed dynamic hysteresis mirrors the static hysteresis on the tested surfaces, suggesting an inverse relationship with molecular flexibility – higher flexibility correlates with lower dynamic hysteresis.

In addition, by attributing the total mechanical work done by the unbalanced Young's force to contact line frictional dissipation, we can evaluate the contact line friction coefficient through both integrated and discrete forms, capturing both overall magnitude and instantaneous variations. The results indicate an absence of noticeable correlation between dynamic hysteresis and the contact line friction coefficient, suggesting that the observed hysteresis loop is not a result of dissipative effects.

To establish suitable boundary conditions for modeling oscillatory contact line behavior, we generalized Hocking's linear law by incorporating the capillary coefficient λ as a function of the time derivative of the logarithm of contact line velocity. For a sinusoidal motion, this expression simplifies into a complex function of frequency ω, validating Miles's idea [23]. Additionally, we assessed the applicability of GNBC and its modified counterpart during oscillatory wetting. While GNBC generated dynamic hysteresis due to the shear term, the predictions proved unrealistic. In contrast, with a proper choice of slip length and contact line friction coefficient, the modified GNBC can predict dynamic hysteresis in an acceptable way.

This study underscores the intricate interplay of dynamic forces and material properties governing oscillatory wetting behavior, paving the way for further exploration and refinement of models describing dynamic wetting phenomena.

**Appendix 1: Discrete form of contact line friction coefficient.**

Figure 11(a) depicts the temporal evolution of the discrete $\mu_{f,i}$. Notably, figure 11(b) reveals a discernible elevation in the contact line friction coefficient as the contact line approaches the positions of maximum displacement, marking the stick-slip region. In figure 11(c), the correlation between $\mu_f$ and $\theta_d$ is illustrated. It is noteworthy that, during variations in the contact angle near its equilibrium value, $\mu_{f,i}$ exhibits higher values, thereby substantiating the applicability of the $\mu_f$ model proposed by Amberg [33] as a function of contact angle. In addition, the plot in panel (d) demonstrates the velocity dependency of the dynamic contact line friction coefficient, $\mu_{f,i}$: when the contact line transiently approaches zero speed, $\mu_{f,i}$ increases exponentially, indicating a slip-to-stick transition.

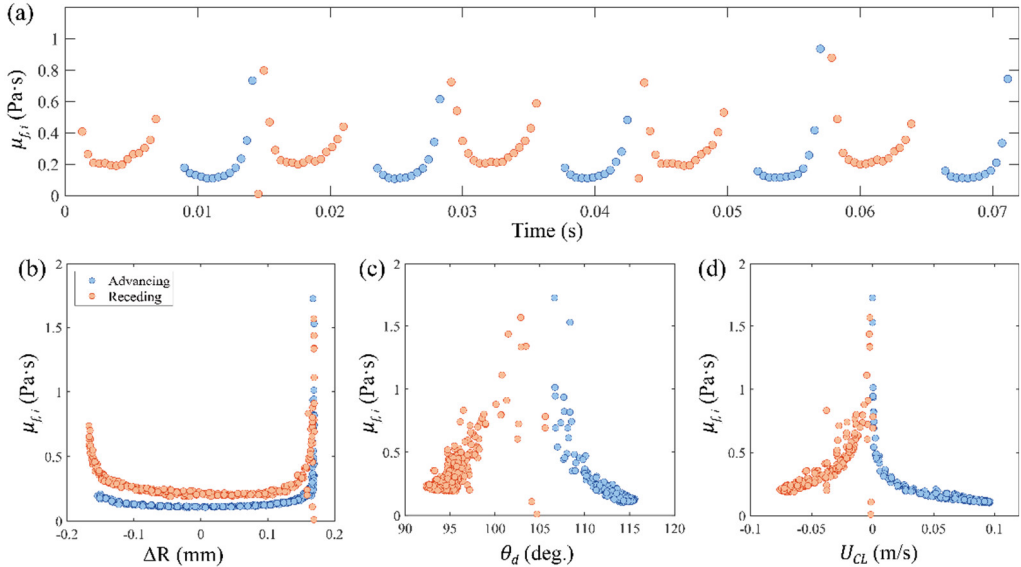

FIGURE 11. Discrete $\mu_f$ distributed with: (a) Time. (b) Contact line displacement. (c) Contact angle. (d) Contact line velocity.

**Appendix 2: The physical interpretation of the experimental contact line friction coefficient.**

The physical interpretation of the experimentally derived contact line friction coefficient merits attention. As noted in the introduction, the dynamic variation of the macroscopic angle arises from diverse contributions across various scales. The hydrodynamic model addresses viscous friction but overlooks the velocity dependence of microscopic angle caused by molecular resistance. In contrast, MKT-based model neglects viscous dissipation while focusing solely on local friction near the contact line at the molecular scale. The $\mu_f$ extracted in our experiment is more likely a collection of both dissipative effects.

The hydrodynamic model expresses the velocity dependence of viscous bending as [9]:

$$g(\theta) - g(\theta_m) = Ca\, ln\left(\frac{L}{L_m}\right), \tag{B 1}$$

where $\theta_m$ and $L_m$ represent the angle and length scale of the microscopic inner region. The model, however, is unable to determine whether $\theta_m$ and $L_m$ vary with velocity. On the other hand, the MKT-based model proposes that the microscopic angle $\theta_m$ directly depends on velocity due to contact line friction from molecular jumping activities. However, this MKT-based model neglects viscous bending, resulting in the macroscopic angle equating to the microscopic angle and becoming velocity-dependent [36].

The experimentally observed variation in the macroscopic angle may encompass effects from both scales, as indicated by a model combining molecular dissipation and viscous resistance [36,43]:

$$\gamma(cos\,\theta_e - cos\,\theta)U = \frac{6\mu}{\theta}ln\left(\frac{L}{L_m}\right)U^2 + \xi U^2. \tag{B 2}$$

Here, $\xi$ represents the coefficient of wetting-line friction interpreted on the molecular scale. Consequently, the experimentally calculated friction coefficient $\mu_f$ can be expressed as the sum of the two components:

$$\mu_f = \frac{6\mu}{\theta}ln\left(\frac{L}{L_m}\right) + \xi. \tag{B 3}$$

Recently, the tapping mode AFM technique has unveiled convex nanobending as a crucial link between molecular-scale and mesoscopic-scale angles [6,8,44]. Chen *et al.* [8] observed that nanobending exhibits velocity dependence. This discovery implies that the experimentally assessed friction coefficient $\mu_f$ should incorporate additional components beyond the right-hand side of Eq. (B 3) to accommodate the contribution from nanobending.

Once again, we emphasize that using a constitutive relation between the dynamic contact angle and contact line speed is one way to describe dynamic contact line behavior. In this approach, the effective contact line friction coefficient is introduced to linearize and simplify the relation. However, other approaches, such as highly resolved direct numerical simulations (DNS) based purely on a hydrodynamic perspective without considering the velocity dependence of the microscopic angle, can also successfully describe dynamic wetting phenomena in complex systems like curtain coating.

**Acknowledgements**

We extend our appreciation to Prof. Shu Takagi and Dr. Timothée Mouterde for their invaluable discussions and kind support in the development of this work. Additionally, we express our gratitude to Michele Pellegrino for having an insightful discussion. This work was partially supported by Japan Society for the Promotion of Science (JSPS) Grant 22H04950. Jiaxing Shen gratefully acknowledges support by SPRING GX project.

**Declaration of interests**

The authors report no conflict of interest.